\documentstyle[aps,prl,floats,epsf]{revtex}
\draft
\tighten
\title{Monte Carlo simulation of a
two-dimensional continuum Coulomb gas}
\author{Jack Lidmar and Mats Wallin}
\address{Department of Theoretical Physics,
Royal Institute of Technology, S-100 44 Stockholm, Sweden}
\date{\today}
\begin{document}
\twocolumn

\maketitle

\begin{abstract}
We study the classical two-dimensional Coulomb gas model for thermal
vortex fluctuations in thin superconducting/superfluid films by Monte
Carlo simulation of a grand canonical vortex ensemble defined on a
continuum. The Kosterlitz-Thouless transition is well understood at
low vortex density, but at high vortex density the nature of the phase
diagram and of the vortex phase transition is less clear. From our
Monte Carlo data we construct phase diagrams for the 2D Coulomb gas
without any restrictions on the vortex density.  For negative vortex
chemical potential (positive vortex core energy) we always find a
Kosterlitz-Thouless transition.  Only if the Coulomb interaction is
supplemented with a short-distance repulsion, a first order transition
line is found, above some positive value of the vortex chemical
potential.
\end{abstract}

\section{Introduction}

Physical systems, which are effectively two dimensional and whose
important thermal excitations are vortices, can undergo a
Kosterlitz-Thouless (KT) transition \cite{KT}.  A prototype model for
such systems is the two-dimensional Coulomb gas (2D CG), and it is
well known to have two distinct phases.  In the low temperature phase
vortices are present only in tightly bound vortex-antivortex pairs
(vortex insulator). In the high temperature phase, above the KT
transition temperature, free vortices are present (vortex metal).
Examples of such physical systems are: thin-film superconductors,
two-dimensional superfluids, Josephson-junction arrays,
two-dimensional melting, and double layer quantum Hall systems,
etc~\cite{RMP,Nienhuis,DLQHS}.  The KT transition has been studied
theoretically in detail, but few rigorous results are established and
some important uncertainties remain.  Notably, how good are the
renormalization group treatments, that are typically justified at low
vortex density, in the high temperature phase, and what is the nature
of the phase transition in the dense limit?  When and how do first
order transitions appear?  In this paper we address these issues by a
grand canonical Monte Carlo simulation of a two dimensional Coulomb
gas model defined on a continuum.

Considerable theoretical understanding of the KT transition has been
gained from various analytic approaches.  The most direct analytic way
to get the KT transition and an approximate phase diagram for the 2D
Coulomb gas is provided by Kosterlitz real-space renormalization group
(RG) equations~\cite{Kosterlitz}.  The equations are justified at
small vortex density, and give a phase diagram containing two phases:
the superfluid phase and the vortex metal phase where superfluidity is
destroyed, separated by the KT transition line.  Kosterlitz equations
can be viewed as the lowest-order equations in an expansion in the
vortex fugacity, $z$, which controls the vortex density (small
fugacity means small density).  Next order RG equations have been
suggested by various authors, e.g.\ Amit et al.~\cite{Amit}, and
recently by Timm~\cite{Timm}.  All these equations coincide with
Kosterlitz equations at low vortex density, where the higher-order
terms are small, and give qualitatively similar phase diagrams, with a
KT transition extending to high vortex density.

However, it is possible that qualitatively new physics appear when the
corrections become big and the small-$z$ expansion is no longer
justified.  Minnhagen has constructed generalized RG
equations~\cite{PM,PM+MW}.  These equations come from a cumulant
expansion, and are of infinite order in $z$ (they resum an infinite
subset of terms).  These equations are justified both in the limits of
high and low fugacity, and in between, at finite fugacity, one may
hence expect them to be better than Kosterlitz-type equations that are
only justified at low $z$.  From these equations, Minnhagen and Wallin
found a qualitatively new phase diagram.  Here the KT transition line
ends at a finite temperature and fugacity, $(T^*,z^*) \approx
(0.144,0.054)$, and below this temperature a first order transition
line replaces the KT line~\cite{PM+MW,Thijssen}.  Several other papers
have also discussed first order transitions~\cite{firstorder,Fisher}.
These examples illustrate that the phase diagrams for the 2D CG from
analytic treatments often come out quite differently at finite vortex
density (see Fig.~\ref{phasediagram} below).  Some methods, but not
all, give first order transitions.

Additional information about the phase diagram of the 2D CG model,
besides the somewhat unsettled picture from analytic calculations, can
be obtained from computer simulations.  Calculations of the phase
diagram of the 2D CG by Monte Carlo simulation have been done by
various authors.  Lee and Teitel \cite{Teitel} have done simulations
of the 2D Coulomb gas defined on square and triangular lattices.
These simulations give a rich phase diagram, containing the expected
KT transition at low vortex density (in agreement with RG theories), a
first order transition at high vortex density, and more.  The first
order transition is from the low-temperature vortex dipole phase into
a dense vortex-antivortex crystal above some critical vortex chemical
potential, with one vortex on each lattice site and thus having the
same lattice structure as the underlying discretization lattice.
These crystal phases have two distinct kinds of order: superfluid
order characterized by a finite macroscopic superfluid density, and
crystalline positional long range order (LRO) with a staggered
Ising-type order parameter for the vorticities.  These two types of
order both survive at finite temperature in the lattice systems, and
they give two distinct transitions as the temperature is increased.
Caillol and Levesque have done simulations of a CG defined on the
surface of a sphere, without any discretization of
space~\cite{Caillol}.  They find indications for first order
transitions at finite vortex density, but they did not present a
finite size scaling analysis, which is desirable in order to
extrapolate the data to the thermodynamic limit.  Similar findings
with later and more accurate methods have been reported recently
\cite{Orkoulas}.

Taken together, analytic results and previous simulations sometimes
give support for first order transitions, besides the usual KT
transition, and sometimes not.  This motivates further work both
analytically and also by new simulations.  In this paper we present
simulations of the 2D Coulomb gas in a grand canonical ensemble with
vortex positions defined on a continuum, without using any underlying
discretization lattice.  The lattice models have important physical
realizations, for examples, in networks of Josephson junction arrays
or granular superconductors, but for homogeneous two dimensional
superfluids and superconductors continuum models are appropriate.
Furhermore, continuum simulations are more similar to RG approaches
that use continuum models.  When are any differences between a lattice
model and a continuum model expected?  Lattice simulations will
correctly give universal critical properties, and are reasonable when
the important length scales are much larger than the lattice constant,
i.e.\ at low densities.  Critical phenomena with a diverging
correlation length should thus be well captured by such lattice
simulations.  However, to investigate general properties of a system
defined on a continuum, such as the form of the phase diagram, and in
particular what happens when the vortex gas becomes dense, one should
use a limit of a small lattice constant, or do simulations directly in
a continuum.  In a continuum system, without an underlying
discretization lattice, we expect the positional LRO of the
vortex-antivortex crystal states found in lattice simulations at high
chemical potential to disappear at any non-zero temperature.  However,
the more relevant question is what happens to the superfluid density:
does it stay finite at any nonzero temperature when the lattice is
removed?

We now summarize our results: Using finite size scaling of our Monte
Carlo data, we construct phase diagrams of the continuum 2D CG.  We
compare these with phase diagrams obtained from several different RG
treatments.  As expected, they all agree and coincide at small vortex
density, where the RG equations become exact.  For negative vortex
chemical potential (positive vortex core energy) we always find a KT
transition.  We find a first order transition line, above some
positive value of the chemical potential, in the case when the Coulomb
interaction is supplemented with a hard core repulsion, but for
soft-core vortices (without any short range repulsion) we do not find
any first order transitions.  In our continuum simulation, we do not
get vortex-antivortex crystal phases, as in the lattice simulations of
Lee and Teitel, with positional LRO or a finite superfluid density
at any of the finite temperatures where we were able to converge our
simulation.  Some of our results have been obtained independently in a
somewhat similar simulation by Holmlund and Minnhagen~\cite{Kenneth}.

The paper is organized as follows: In Sec.\ II the definition of the
Coulomb gas is introduced and some previous results are discussed in
some detail.  Sec.\ III describes our Monte Carlo calculation.  Sec.\
IV contains our results, and Sec.\ V contains discussion of our
results and conclusions.

\section{The two-dimensional Coulomb gas model}

In this section we will describe the definition and some details of
the two dimensional Coulomb gas (2D CG) model.  This model follows in
certain limits from the Ginzburg-Landau (GL) theory, namely when the
only important fluctuations in the GL order parameter field $\Psi{(\bf
r})=|\Psi{(\bf r})|\exp(i\phi({\bf r}))$ are phase fluctuations, which
leads to logarithmically interacting vortices and antivortices,
corresponding to positive and negative ``charges'' in the 2D Coulomb gas.

The 2D Coulomb gas model
is defined by the grand canonical partition function
\begin{equation}
Z = \sum_{N=0,2,4, \dots ,\infty}
    \frac{1}{N_+! N_-!}
    \left( \prod_{j=1}^N \int \frac{d^2 r_j}{\zeta} \right)
    e^{-\beta(H - \mu N)},
\label{Z}
\end{equation}
where $N_+, N_-$ is the number of positive and negative
Coulomb gas particles (i.e., vortices and antivortices),
and $N=N_+ + N_-$ is the total number of particles.
We will only consider the neutral Coulomb gas, where $N_+=N_-$, which
corresponds to no external magnetic field applied to the
superconductor, or no net rotation of the superfluid.
The dimensionless inverse vortex temperature is
$\beta = 1/T = 2 \pi \rho_0 \hbar^2/m^* k_B T^{physical}$, and
$\mu = -E_c$ is the chemical potential of the CG particles, $E_c$
being the vortex core energy. $\rho_0$ is the superfluid density in
the absence of vortices, and $m^*$ is the mass of the boson
responsible for superfluidity/superconductivity.
The phase-space division $\zeta$ is an arbitrary constant
which we will set to $\zeta=1$.

The Hamiltonian is
\begin{equation}
H = \frac{1}{2} \int q({\bf r})G({\bf r}-{\bf r'})
q({\bf r'}) d^2{\bf r}d^2{\bf r'},
\end{equation}
where $q({\bf r}) = \sum_i s_i f({\bf r}-{\bf r}_i)$ is the vortex
density, \mbox{$s_i=\pm 1$} is the vorticity or CG charge of the
particle at ${\bf r}_i$ and $G$ is the solution to Poisson's equation,
$\nabla^2 G({\bf r}) = - 2\pi\delta({\bf r})$.  For a 2D infinite
system this gives $G({\bf r}) \sim - \ln |{\bf r}|$.  The function
$f({\bf r})$ is the charge distribution for a single particle, which
will be discussed in more detail below.
Defining
\[
V({\bf r}) = \int f({\bf r}')G({\bf r}+{\bf r}'-{\bf r}'')f({\bf
r}'')d^2{\bf r}'d^2{\bf r}'',
\]
the Hamiltonian can be written
\begin{eqnarray}
H &=& \frac{1}{2} \sum_{i,j} s_i V({\bf r}_i-{\bf r}_j) s_j \\
&=& \frac{1}{2} \sum_{i \ne j} s_i
\left[ V({\bf r}_i-{\bf r}_j) - V(0)\right] s_j +
\frac{1}{2} V(0)\left(\sum_i s_i\right)^2.	\label{H} \nonumber
\end{eqnarray}
For an infinite system (or a finite system with periodic boundary
conditions) $V(0)$ is actually infinite, forcing $\sum_i s_i = 0$ to
make the last term vanish (i.e.\ the system must be neutral).

The interaction has to be regularized at short distance or otherwise
the logarithmic divergence at $r=0$ will make the system unstable.
This is usually done by putting the system on a lattice, but here we
are working on a continuum and we thus have to modify the interaction.
This is done by defining a (normalized) ``charge distribution''
$f({\bf r})$ of a vortex, here taken as a Gaussian
$f({\bf r}) = \frac{1}{\pi r_c^2}\exp(-{\bf r}^2/r_c^2)$, where $r_c$
is a measure of the vortex core radius.  The physical origin of a
short distance cutoff in superfluids comes from the fact that the
current must be finite at the vortex center.  The charge distribution
describes how the magnitude of $\Psi$ is suppressed to zero in the
vortex core.

In addition to having a finite ``charge distribution'' we will
sometimes treat the vortices as hard disks, thus excluding the
possibility of overlapping vortex configurations.  This case is
convenient to simulate because it leaves no ambiguity in keeping track
of the number $N$ of vortices in the system.  We will also consider
the case of soft cores where vortex cores are allowed to overlap.  In
this case the number of vortices $N$ in the system is not well defined
in the case of many overlapping vortex cores, and this is a serious
complication because $N$ is explicitly needed in the evaluation of the
Boltzmann factors, $e^{-\beta(H-\mu N)}$ in the simulation.  We
neglect this difficulty and just take the number of vortices as the
number inserted into the system, whether or not they overlap.  To
overcome this simplification one should instead consider a
Ginzburg-Landau model, but this is beyond the scope of this paper.  A
hard core repulsion makes it possible to study the model for positive
chemical potentials, i.e.\ for negative core energies.  In particular
a repulsion and a negative core energy are necessary requirements for
observing vortex-antivortex crystal phases.  We will return in the
final section to a comparison between the hard and soft disk cases,
and a discussion of possible physical realizations.

In principle the relation between the hard core diameter and the width
of the charge distribution $r_c$ is a tunable parameter, and it has
quantitative effects.  In particular, the location of the KT line in
the phase diagram will depend strongly on this choice.  A natural
choice for the width of the charge distribution, $r_c$, is that which
gives an interaction which in an infinite system behaves
asymptotically as $V({\bf r})= -\ln|{\bf r}|$ for large $|{\bf r}|$.
This happens for $r_c = \frac{1}{\sqrt{2}}e^{\gamma/2} \approx
0.9437$, where $\gamma$ is Euler's constant.  This choice simplifies
comparison with analytic results, since they usually also assume a
pure logarithm at large distance.  Other choices of $r_c$ lead to a
constant shift in $V$ at large distances.  We did some limited
calculations for other choices of $r_c$ to check the quantitative
effects on the phase diagrams, and some results for the choice
$r_c=0.1$ will be given below.  The hard core diameter is arbitrarily
set to one (or zero in the soft core case).

In simulations of the 2D Coulomb gas model we are restricted to finite
system sizes, and in order to mimic the thermodynamic limit we use
periodic boundary conditions, as usual.  The long-range logarithmic
vortex interaction has to be modified for this case, and the easiest
way to accomplish this is by expanding $V$ in a Fourier series
(taking into account the periodic boundary conditions and the finite
charge distribution):
\begin{equation}
V({\bf r}) - V({\bf 0}) = \frac{1}{L^2}\sum_{\bf k}
\frac{2\pi}{{\bf k}^2}|f_{\bf k}|^2
\left(e^{i {\bf k} \cdot {\bf r}} - 1\right).
\label{V}
\end{equation}
Here $f_{\bf k} = \exp(-{\bf k}^2 r_c^2/4)$ is the Fourier transform
of the ``charge distribution'' $f({\bf r})$ and $L$ is the linear
system size.
The allowed wave vectors are quantized by the periodic boundary
conditions and given by ${\bf k}=\frac{2\pi}{L}\left(n_x,n_y\right)$,
where $n_x, n_y$ are integers.
This makes the interaction periodic with period $L$.

The Gaussian cutoff of the interaction is convenient when evaluating
the interaction since it means only a relatively small number of wave
vectors in Eq.\ (\ref{V}) will contribute.  To get a quick evaluation
of the interaction in continuum simulations we use a look-up table
defined on a fine lattice and a bilinear interpolation to the vortex
positions.  We will only consider relatively small systems, and do
therefore not use further convergence accelerations for evaluating the
Coulomb potential.

There are two related main methods to detect a KT transition from
physical measurements involving vortex dynamics.  The first one is to
look for the universal jump in the superfluid density, and the other
is the universal nonlinear voltage-current characteristic $V \sim J^3$
at $T=T_c$.  In the Coulomb gas the universal jump is seen in the
dielectric response function, whose inverse is given by
\begin{equation}
\epsilon^{-1}({\bf k}) = 1 -
\frac{V({\bf k})}{L^2T}\left< q_{\bf k} q_{-\bf k} \right>,
\end{equation}
where $V({\bf k})$ and $q_{\bf k}$ are the Fourier transforms of the
interaction and charge density, respectively.  For a superfluid,
$\epsilon^{-1}({\bf k \rightarrow 0})$ is proportional to the
macroscopic superfluid density $\rho_s$, fully renormalized by vortex
excitations.  This corresponds to the macroscopic spin stiffness in
the XY model.
The universal jump at the KT
transition is given by \cite{univjump}
\begin{equation}
\frac{1}{\epsilon(0) T_c} =
\left\{ \begin{array}{ll}
          4 & \mbox{at $T=T_c^-$} \\
          0 & \mbox{at $T=T_c^+$}
\end{array}
\right.
\label{univjump}
\end{equation}
In simulations,
the inverse dielectric constant is usually defined only at finite
wave vectors in the Coulomb gas.  To use this quantity in a
simulation requires extrapolation from the smallest nonzero wave vector
$k=2 \pi/L$ to zero.  An alternative is to modify the definition of
the model to allow for zero wave vector excitations corresponding to
vortex currents across the system \cite{Peter,Vallat}.  This is
accomplished by adding to the Hamiltonian the term
\begin{equation}
H' = \frac{\pi}{L^2}{\bf P}^2,
\end{equation}
where the polarization is ${\bf P} = \sum_i s_i {\bf r}_i$.
The ${\bf k = 0}$ response is now given by
\begin{equation}
\epsilon^{-1}=1-\frac{\pi}{L^2T}\left<{\bf P}^2\right>.
\label{eps0}
\end{equation}
In the lattice version of the model this quantity corresponds exactly
to the spin stiffness (helicity modulus) of the 2D XY model after
replacing the cosine interaction with the Villain interaction
\cite{Peter,Vallat}.
For finite system sizes there will be a logarithmic correction at $T_c$
given by \cite{correction}
\begin{equation}
\epsilon^{-1}(T_c,L) = \epsilon^{-1}_\infty
  \left( 1 + \frac{1}{2\log L + C} \right).
\label{correction}
\end{equation}
We will use this equation below to locate the KT transition
temperature from Monte Carlo data on finite systems.
We now turn to the simulation methods used in our calculations.

\section{Monte Carlo methods}

In this section we describe our Monte Carlo (MC) algorithm in some
detail.  Our algorithm simulates a grand canonical ensemble of
particles on a continuum, using a finite step length for the various
MC moves.

Most of the previous simulations of the 2D CG were done on the lattice
version of the model.  The lattice simulation is easily implemented by
considering one single type of MC trial moves: adding
vorticity-neutral pairs on randomly selected nearest neighbor pairs of
lattice sites.  If the energy change of inserting the vortex pair is
$\Delta E$ the attempt is accepted according to the usual Metropolis
algorithm with probability $\exp(- \beta \Delta E)$. This algorithm is
effective in order to construct the equilibrium phase diagram as a
function of temperature $T$ and vortex chemical potential $\mu$. This
simulation accurately verifies that the Coulomb gas has a KT
transition at low vortex density in full agreement with analytic
results~\cite{Teitel}. At high vortex density, however, the simulation
becomes sensitive to the discretization of space, and in order to
investigate the properties of the model in this limit we do our
simulations on a continuum.

Below we will describe our Monte Carlo algorithm, which is an
extension of an algorithm described by Valleau et al.~\cite{Valleau}.
In a grand canonical Monte Carlo simulation the number of particles is
fluctuating.  Thus Monte Carlo moves which change the particle number
must be considered.  There are many ways to accomplish this, all
having in common that the acceptance probabilities must be different
from the usual $e^{-\beta\Delta E}$ in order to ensure detailed
balance.  In our case (periodic boundary conditions and no external
magnetic field) we must also require that the particles are created
and destroyed in pairs of opposite charge, to keep the system neutral.
Since the Coulomb interaction favors configurations of particles in
tightly bound pairs, we find it convenient to use an algorithm in
which the particles are attempted to be placed close to each other.
This reduces the correlation time of the MC simulation considerably
compared to the case where particles are created at independent random
positions.  This helps to obtain reasonably fast convergence of the
simulation.

The attempted Monte Carlo moves are:
1.\ Creation: A particle is inserted randomly in the system, and then
another particle with opposite charge is inserted 
at random within a distance $d$ from the first one.
2.\ Destruction: A randomly chosen particle, and a particle of
opposite charge within a distance $d$ from the first one (if any) is
deleted from the system.
3.\ Movements of particles: A particle (or a pair of particles) is
moved a random distance.
The value of $d$ can be tuned to optimize convergence.
The acceptance probabilities for these moves can be found from the
following argument.

The probability of a state $i$ in the grand canonical ensemble is
given by
\begin{equation}
P_i = \frac{1}{Z}e^{-\beta(E_i - \mu N_i)}.
\end{equation}
Let $w_{ij}$ denote the transition probability to go from state $i$ to
state $j$ in the Markov chain.  A sufficient condition that the
probability distribution will converge towards the equilibrium one
(besides ergodicity) is given by the condition of detailed balance:
$w_{ij} P_i = w_{ji} P_j$.  Now let $t_{ij}$ be the transition
probability for the {\em trial} moves between states $i$ and $j$
(i.e.\ the conditional probability to attempt to go to state $j$ given
that the current state is $i$), and $a_{ij}$ the probability that the
corresponding trial move gets {\em accepted}.  This means that
\begin{equation}
w_{ij} = t_{ij} a_{ij} \qquad(i \ne j),
\qquad \qquad w_{ii} = 1 - \sum_{j \ne i} w_{ij}.
\end{equation}
The trial transition probability for creating a pair is equal to the
inverse of the volume of the phase space in which the new particles
are attempted to be placed (normalized by the phase space division),
while for the destruction moves described above it is equal to the
inverse of the total number of ways to remove a pair of particles.
Both these quantities are needed for the evaluation of the acceptance
probabilities in both the creation and destruction moves. If we
consider the transitions between states $i$ and $j$, with total number
of particles $N_j=N_i+2$ we have
\begin{equation}
t_{ij} = \frac{\zeta^2}{V \Omega }, \qquad
t_{ji} = \frac{1}{(N_j/2) N_\Omega},
\end{equation}
where $V = L^2$ is the (2D) volume of the system (the phase space
available for the first particle created) and $\Omega$ is the phase
space available for the second particle, given by $\Omega=\pi d^2$ for
soft core particles, and in the case of hard cores by
$\Omega=\pi(d^2-1)$ (the excluded volume due to the hard disk diameter
$1$ is subtracted).  $N_j/2$ is just the number of ways to
choose the first particle of a given charge in a destruction move.
The number of ways to choose the second one is given by $N_\Omega$,
which we define as the number of oppositely charged particles within a
distance $d$ from the first particle chosen in a pair to be destroyed
(or from the first one of a hypothetically created pair).  Detailed
balance now follows if the acceptance probabilities are chosen so that
\begin{equation}
\frac{a_{ij}}{a_{ji}} = \frac{t_{ji}}{t_{ij}} \frac{P_j}{P_i} =
\frac{V \Omega}{(N_j/2)N_\Omega} e^{-\beta( E_j-E_i - 2\mu) - 2\ln(\zeta)}.
\end{equation}
We see that the precise value of the phase space division $\zeta$ only
enters as a shift in the chemical potential. In the following we will
set it equal to one. Thus we choose the acceptance probability for
creations to be $\min(1,a_{ij}/a_{ji})$, and for destructions
$\min(1,a_{ji}/a_{ij})$, while it is the ordinary $e^{-\beta\Delta E}$
for displacements.  The value of $d$ is arbitrarily chosen to 2.  The
creations and destructions must be attempted with equal probability.

A MC sweep consists of $L^2$ trial moves.  At each temperature about
$10^4$ initial sweeps were discarded to equilibrate the system, and
then between $10^5$ to $10^6$ were used to calculate averages.

\section{Results}

\begin{figure}
\epsfxsize=9truecm\epsffile{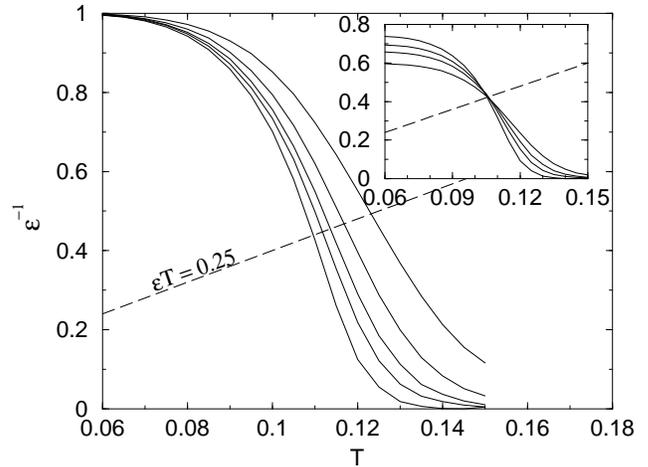}
\caption{
Location of the KT temperature (in dimensionless units) from MC data for
the dielectric function $1/\epsilon(0)$ of the CG.
System sizes are $L=6,8,10,12,16$ from top to bottom.
The chemical potential is $\mu=-0.16$.
Inset shows MC data corrected for finite size effects
according to the text, making curves cross at $T_c$
given by the universal jump criterion (dashed line).
The smallest system ($6 \times 6$) deviates slightly
and was discarded.
}
\label{eps}
\end{figure}

We now turn to our results.  Below we will construct phase diagrams
for the 2D CG model.  To do this we must first locate the phase
transition points from Monte Carlo data.  Figure~\ref{eps} shows how
to determine the KT transition temperature from MC data for the
dielectric function $1/\epsilon(0)$ given by Eq.\ (\ref{eps0}).  The
data in the figure is for vortex chemical potential $\mu=-0.16$ and
system sizes are $L \times L$ with $L=6,8,10,12,16$ ordered from top
to bottom.  The curves are actually formed from straight line segments
interpolating between MC data points.  Since the statistical errors
are small these straight line segments in this figure form quite
smooth curves and we do therefore not show the data points or error
bars.  The straight dashed line is the universal jump condition given
by Eq.\ (\ref{univjump}), and $T_c$ would be where an infinite system
crosses this line.

The inset shows how to construct this intersection point by including
a correction to the scaling formula, given by Eq.\ (\ref{correction}).
This practically eliminates finite size effects at the critical point,
which is at the common crossing of the curves for different sizes with
the straight universal-jump line.  The smallest system size ($6 \times
6$) deviates slightly and was left out of the fit.  This procedure
gives a quantitatively rather accurate extrapolation to the
thermodynamic limit allowing us to determine $T_c \approx 0.105$ at
this value of $\mu$.  Similar calculations at other values of $\mu$
allow us to construct the main parts of the phase diagram of the
model.

\begin{figure}
\epsfxsize=9truecm\epsffile{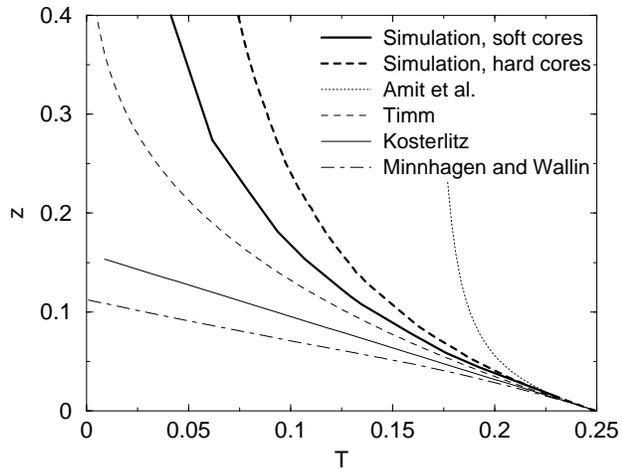}
\caption{
Phase diagram of the 2D Coulomb gas.
The various curves are transition lines determined by different methods
(see text).
Note that all curves coincide close to the point $(T,z)=(1/4,0)$
in the phase diagram, but deviate at finite $z$.
The bottom curve is a KT line for $T>0.144$ and
first order for $T<0.144$.  All other curves are KT
transition lines.
}
\label{phasediagram}
\end{figure}

The main issue of this paper is to investigate the nature of the phase
diagram of the 2D CG at finite vortex densities, and to compare with
qualitative and quantitative features of existing theories.
Figure~\ref{phasediagram} shows the phase diagram of the 2D CG for low
to moderate vortex densities.  The curves in the figure are various
transition lines computed from our continuum simulation for the case
of hard and of soft cores, lowest order equations by
Kosterlitz~\cite{Kosterlitz}, next order RG equations by Amit et
al.~\cite{Amit} and by Timm~\cite{Timm}, and the equations of
Minnhagen \cite{PM+MW}.  All curves are KT transition
lines except the low-$T$ part of the bottom curve which is a first
order line; the KT part of the bottom curve extends between $T \approx
0.144$ and $T=1/4$.  We see that the KT lines from all determinations
coincide close to the point $(T,z)=(1/4,0)$ in the phase diagram, but
otherwise they deviate significantly from each other.  The agreement
is expected because close to this point all theories are exact by
construction, and this shows that the simulation works as expected.
It is, however, clearly seen from the figure that the parameter regime
in which the approximate theories give quantitative agreement is very
limited.

Next we will investigate the nature of the phase transition in the
region of high vortex density in the phase diagram.  Here different
things happen for hard and soft disks: hard disk vortices have a
change of ground state when the vortex core energy becomes negative
from an empty system into a square vortex-antivortex crystal.
Soft-core vortices instead change into a normal state where
overlapping cores cover the whole system.  We will here assume
hard-disk vortices.

\begin{figure}
\epsfxsize=9truecm\epsffile{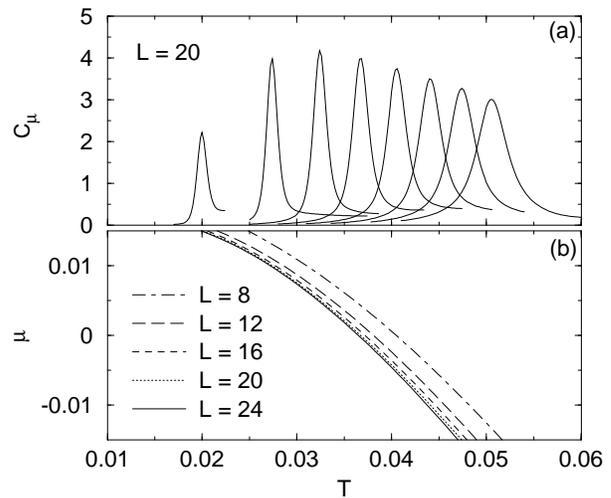}
\caption{
(a) Specific heat as a function of temperature $T$ at fixed chemical
potentials, $\mu = 0.015, 0.01, \ldots , -0.02$, from left to right,
for a system of hard core vortices.  The system size is $20\times 20$.
The curves are formed using the multiple histogram method
\protect\cite{multihist} from three simulations at the parameters
$T=0.0368, 0.0324, 0.0275$ and $\mu = 0, 0.005, 0.01$, respectively.
(b) Location of the peak in the specific heat as a function of $T$,
for several system sizes.
}
\label{cv}
\end{figure}

A first indication that something changes at high densities is given
in Fig.~\ref{cv}(a).  Here we plot the specific heat as a function of
temperature for a number of fixed chemical potentials.  If the phase
transition is of the KT type, there is no divergence
in the specific heat at $T_c(\mu)$, but there is a peak at a somewhat
higher temperature~\cite{Barber}.  In the figure we see that, as the
chemical potential is increased, the peak moves to lower temperature
and gets sharper and higher, until it reaches a maximum around $T
\approx 0.03$, where it starts to decrease.  A possible explanation of
this change of behavior would be that the KT-transition is replaced by
a first order transition line at lower temperature, ending at a
critical point with a diverging specific heat.  Indeed the peak in the
specific heat show strong finite size effects in the whole parameter
region depicted in the figure.  Further support of this interpretation
is found using the histogram methods described below.  It is difficult
to give an accurate estimate of the critical exponents, since the we
do not know exactly where the critical point is.  Our results are,
however, consistent with a value of roughly $\alpha/\nu \approx 1.6$.

\begin{figure}
\epsfxsize=9truecm\epsffile{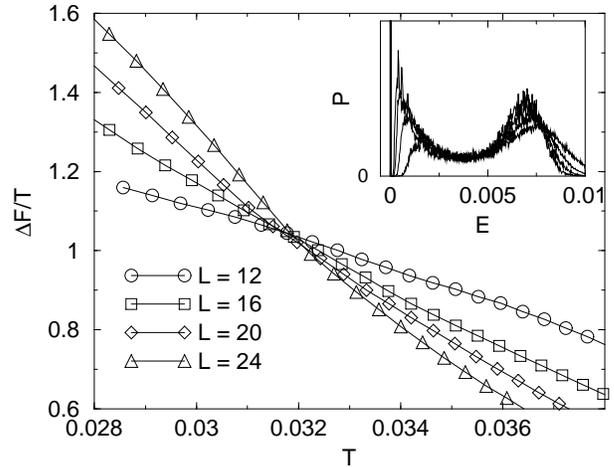}
\caption{
Free energy barrier $\Delta F$ determined from histograms of the
Boltzmann factors in the simulation of vortices with hard cores.
Inset: histograms of the Boltzmann factors $P$ vs.\ energy $E$ at
$\mu=0.008, T \approx 0.029$.  The figure shows three different
behaviors: At low temperature $\Delta F$ grows with increasing system
size indicating a first order transition line, ending at a critical
point at $T \approx 0.032$ where all curves cross.  Above this
temperature the free energy barrier decreases with increasing system
size.  The first order line at low temperature replaces the
KT transition line which starts at $T \approx 0.032$
and goes up to $T=1/4$.
}
\label{deltaf}
\end{figure}

The location of the peak in the specific heat for the hard core case
for different fixed temperatures and system
sizes forms the curves in the $(T,\mu)$-plane shown in
Fig.~\ref{cv}(b).  
Close to these curves we do a more detailed analysis.
We use standard histogram methods, together with a smoothing
procedure described below.  The procedure to locate parameter points
that will extrapolate to the critical parameters in the thermodynamic
limit uses histograms in two related ways.  First, we use the multiple
histogram method \cite{multihist}.  This enables us to extrapolate
thermodynamic quantities to nearby values of $\mu$ and $T$, which
avoids expensive simulations at every point.  Secondly, we use
histograms to distinguish between a continuous and a first order
transition \cite{histogram}.
We constructed histograms using a very small bin size. This produces
very noisy histograms which were then smoothed by forming averages
over the 16 nearest bins.  Close to a first order transition the
probability distribution $P(E)$ will have a double peak structure with
one peak in each phase, separated by a minimum.  An example can be
seen in the inset in Fig.~\ref{deltaf}.  The free energy barrier
between the phases is defined by $\Delta F /T = -
\log(P_{max}/P_{min})$, where $P_{max}$ is the value of the
probability distribution at the peak, and $P_{min}$ is at the valley
in between.  If the barrier increases with increasing system size this
signals that there will be a first order transition in the
thermodynamic limit.  If it decreases there will be no phase
transition, and if it is constant there will be a critical point.
This analysis is usually carried out for parameters at which the two
peaks have equal height.  In our case, however, the histograms are
highly asymmetric, which makes this criterion rather useless.  Instead
we use the following procedure: For a given value of the chemical
potential, we adjust the temperature to the point where the two peaks
have equal {\em weight}.  We accomplish this by repeatedly smoothing
the histogram curve by averaging over neighboring bins until the peaks
become symmetric, and then we adjust $\mu$ and $T$ until their heights
become equal. This method gives a practical way to approximately
locate the parameters at the transition.  We find that for a given
$\mu$ and system size $L$, the points where the two peaks have equal
weight form a curve in the $(T,\mu)$-plane, which depends slightly on
the system size.  In our case this curve coincides to a very good
approximation with the location of the peak in the specific heat (see
Fig.~\ref{cv}(b)), which was used as a preliminary estimate of the
location of the critical parameters for each system size.

Fig.~\ref{deltaf} shows the result of the histogram analysis, for the
case of hard disks.  The free energy barrier $\Delta F$ was
constructed from histograms of the energy distribution in the
simulation.  The inset shows typical such histograms of the Boltzmann
factors $P$ vs.\ energy $E$. 
Three regions can clearly be seen.  At low
temperature, $\Delta F$ grows with increasing system size indicating a
first order transition line ending at a critical point at 
$(T,\mu) \approx (0.032,0.004)$ where all curves cross.  Above this
temperature, the free energy barrier decreases with increasing system
size.  The first order line at low temperature replaces the
KT transition line which starts at $T
\approx 0.032$ and goes up to $T =1/4$.  At the first order transition
there is a discontinuity in the particle density, and in the energy.
None of the phases have any long range positional order.  For soft
disks we have not found any evidence for a (finite temperature) first
order transition.

Simulations on discrete lattices give qualitatively different results
in the high density regime.  Here one finds vortex-antivortex crystals
at finite temperature and high chemical potential~\cite{Teitel}.
These phases have two distinct kinds of order: positional LRO of a
staggered vorticity order parameter commensurate with the underlying
discretization lattice, and finite macroscopic dielectric function.
Associated with these are two separate finite temperature phase
transitions: one Ising type transition and a KT transition.  The Ising
LRO is expected to disappear at any finite temperature in the
continuum limit, when there is no pinning to an underlying lattice,
but both algebraic quasi-LRO and a finite dielectric response function
can in principle remain at finite $T$.  To further investigate the
relation between lattice and continuum simulations, and to see what
happens to the phase transitions found in the lattice Coulomb gas when
the continuum limit is approached, we did a sequence of lattice simulations
with decreasing lattice constants.

\begin{figure*}
\hbox{
\epsfxsize=9truecm\epsffile{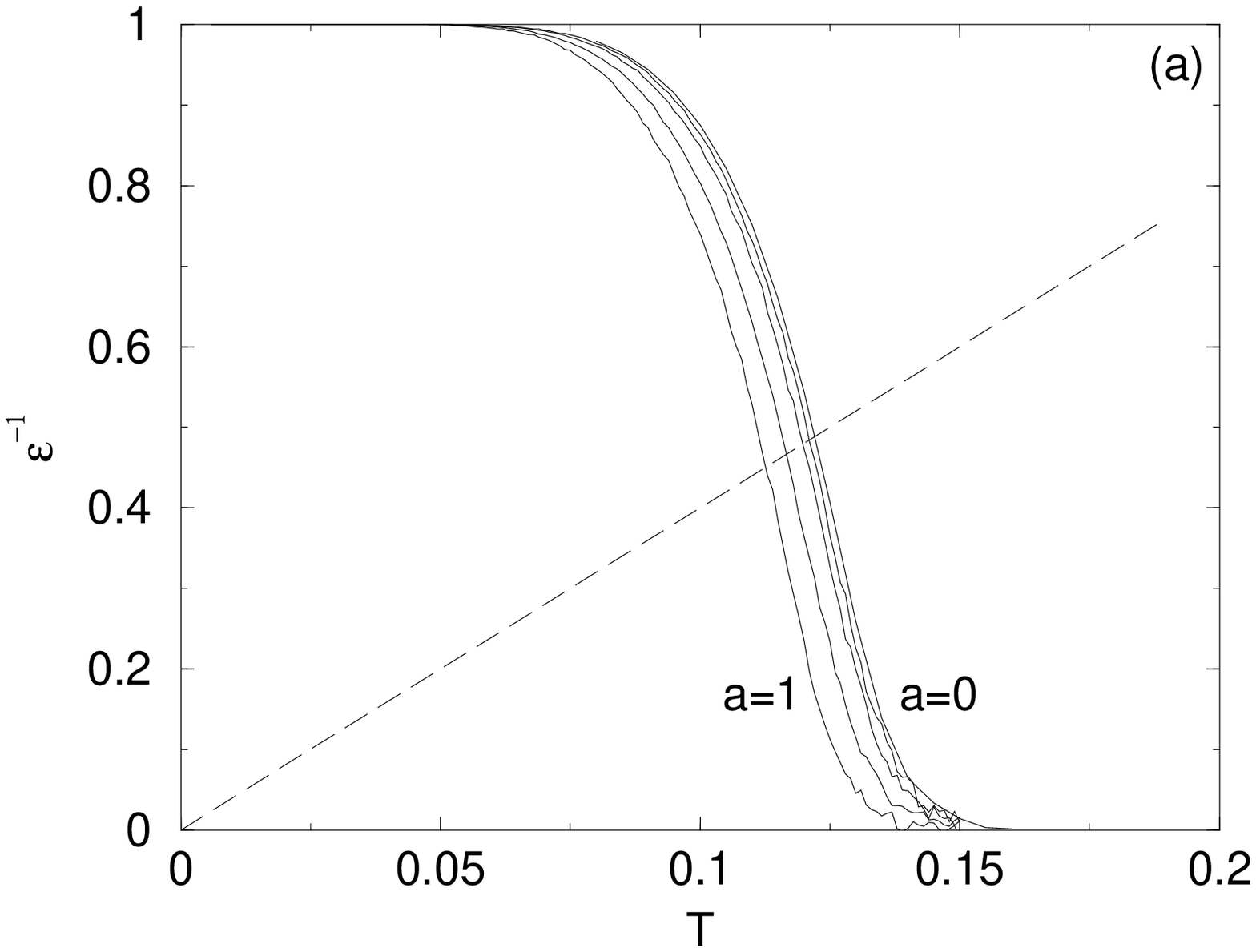}
\epsfxsize=9truecm\epsffile{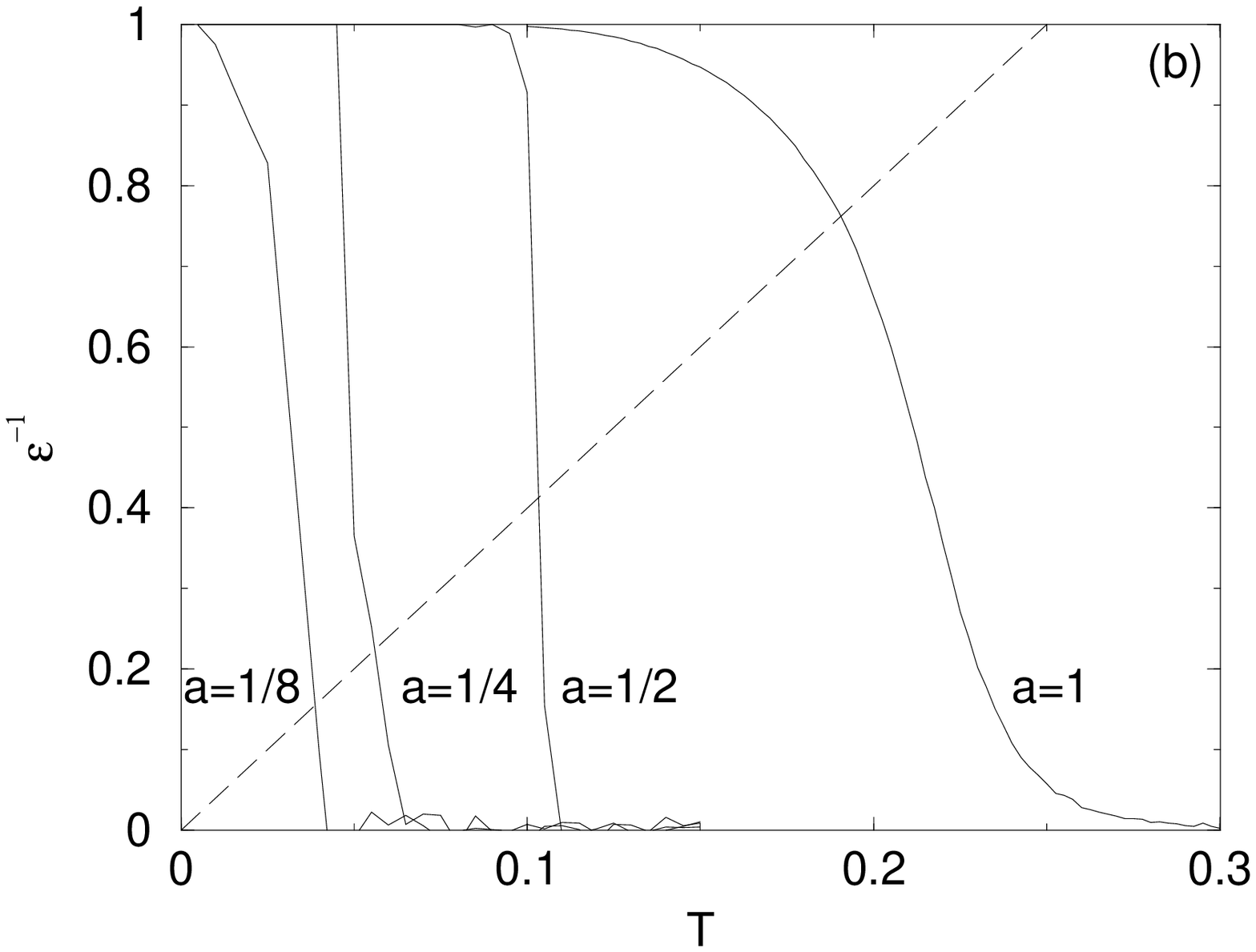}
}
\caption{
Inverse dielectric constant from MC data on lattices with a sequence
of decreasing lattice constants 
$a=1, \frac{1}{2}, \frac{1}{4}, \frac{1}{8}, 0$,
where $a=0$
means continuum simulation.  In (a) the chemical potential is
negative, $\mu=-0.2$, and the KT transition temperature goes up
somewhat when the lattice is removed.  The system size is $16 \times
16$.  In (b) the chemical potential is positive, $\mu=+1.0$, and the
width of the charge distribution is chosen as $r_c = 0.1$ in units of
the hard disk cutoff diameter.  The system size is $8 \times 8$.  Here
the KT transition temperature drops sharply as the discretization
lattice is refined, and any finite dielectric response function of a
vortex-antivortex phase is unobservable at finite temperature in the
continuum case.  The dashed line is the universal jump criterion for
the KT-transition $\epsilon^{-1} = 4T_c$.  The crossing points give
estimates of the transition temperatures for the different curves.  }
\label{lat}
\end{figure*}

Figure \ref{lat} shows data for the inverse dielectric constant
$1/\epsilon(0)$ for lattice constants 
$a=1, \frac{1}{2}, \frac{1}{4}, \frac{1}{8}, 0$, where
$a=0$ means continuum simulation.  In (a) the system size is $16
\times 16$ and the chemical potential is negative, $\mu=-0.2$ which is
in the region where the ground state is empty.  We first observe that
the results of the simulations interpolate smoothly from the lattice
case to the continuum case as the lattice constant is decreased.  This
serves as a useful test that the two different Monte Carlo programs
for the two cases both converge.  The data in the figure imply that
the KT transition temperature goes up somewhat when the discretization
lattice is removed for this value of the chemical potential.  To
investigate why $T_c$ goes up when the lattice is removed, we looked
at the vortex densities.  For the same parameters, the lattice case
has somewhat higher vortex density than the continuum case. This is
because the pinning potential of the lattice favors tightly bound
vortex-antivortex pairs, which leaves more space in the system for
other vortices.  In (b) the chemical potential is positive,
$\mu=+1.0$, for which the corresponding ($T=0$) ground state is a
vortex-antivortex crystal.  Here the width of the charge
distribution is chosen as $r_c=0.1$ in units of the hard disk cutoff
diameter and the system size is $8\times 8$ (this choice of $r_c$
increases the tendency of the vortices to order in a vortex-antivortex
crystal compared to our previous choice $r_c
=\frac{1}{\sqrt{2}}e^{\gamma/2}$).  Here the KT transition temperature
drops sharply as the discretization lattice is refined, and any finite
dielectric function of a vortex-antivortex phase is unobservable at
finite temperature in the continuum case.  Also the transition
temperature for the Ising-Type staggered order, which for lattice
constant $a=1$ is higher than the KT transition, approaches zero as
the lattice constant is decreased.  This demonstrates that the finite
temperature crystal phases shrink down to zero temperature (or at
least to smaller temperature than we could converge numerically) in
the continuum limit.

\section{Discussion}

We have constructed phase diagrams of the 2D continuum Coulomb gas
from grand canonical Monte Carlo simulations, and compared it with
various (approximate) analytical results (see
Fig.~\ref{phasediagram}).  We find that all results agree at small
fugacity, as expected, but at finite fugacity various results differ
significantly both quantitatively and qualitatively.

We studied two different cases, namely vortices with and without an
additional hard disk repulsion.  In the case when the interaction is
purely Coulombic we find that the transition remains continuous,
consistent with the KT theory, at any negative value of the chemical
potential (corresponding to fugacities $z=e^{\mu/T} < 1$).  At $\mu=0$
the system goes normal and the CG model breaks down.  On the other
hand, when a hard core repulsive interaction is added to the Coulomb
interaction, we find evidence from finite size scaling of the Monte
Carlo data for a first order transition at finite but low temperature,
$T \lesssim 0.032$, and at a small positive value of the chemical
potential.
Our critical endpoint $T \approx 0.032$ is roughly
consistent with the one found in Ref.~\cite{Orkoulas}.
We do not find first order transitions at negative
chemical potentials as in Refs.~\cite{PM+MW,firstorder}.

The existence of our first order transition is related to the change
of ground state at the point where the chemical potential exceeds the
Coulomb energy per particle in a square vortex-antivortex lattice,
with lattice constant equal to the hard disk diameter.  This energy
should be compared to the basic energy scale for creating vortices,
the pair creation energy, which is the total energy of a
vortex-antivortex pair whose hard disks touch.  Both the
crystallization energy and the pair creation energy depend very
strongly on the ratio between the hard disk diameter and the short
distance cutoff, $r_c$, in the Coulomb interaction.  Indeed we find
that, in all cases we tried, the value of $\mu$ at which the first
order transition takes place is slightly above the ground state
crystalization energy (but below the pair creation energy).  It is
interesting to note, however, that this finite temperature first order
transition takes place not between the empty state and the
vortex-antivortex crystal, but between a gas-like low density phase
and a liquid-like high density phase, with no long range order.  This
is in contrast with simulations done on an underlying discretization
lattice, where the first order transition is to a vortex-antivortex
crystal~\cite{Teitel}.

To investigate this effect further we did a sequence of lattice
simulations with different lattice constants, $a=1, \frac{1}{2},
\frac{1}{4}, \frac{1}{8}, 0$.  We find that the extent of the LRO
vortex-antivortex crystal phase shrink rapidly towards zero
temperature with decreasing lattice constant, leaving no sign in the
continuum case, at least for the parameters considered by us.  We can,
however, not rule out the possibility of quasi-LRO, orientational
order, etc., at lower temperatures than could be safely converged
in our simulations.  It should be mentioned that in
this parameter regime (high density, low temperature) it is very hard
to converge the simulations, and all our results are for rather small
systems due to limitations in what sizes we are able to converge
safely.

\begin{figure}
\epsfxsize=9truecm\epsffile{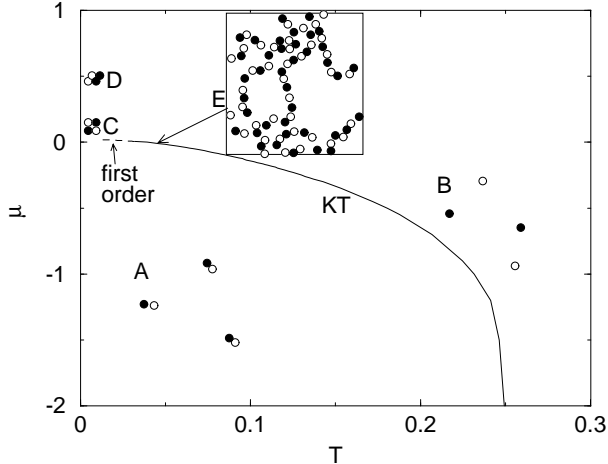}
\caption{
Phase diagram with snapshots of typical configurations inserted.  The
displayed snapshots are the following: (A) superfluid vortex-dipole
phase at low density.  (B) non-superfluid free vortex phase.  (C)
square vortex-antivortex crystal ground state at small
positive vortex chemical potential (negative core energy).  (D)
triangular (frustrated) vortex crystal above some threshold
value of the chemical potential.  (E) characteristic 1D
vortex-antivortex patterns in the regime where the dense vortex liquid
evaporates.
}
\label{snap}
\end{figure}

To summarize our results, we will now discuss several qualitative
features of the phase diagram of the 2D CG with hard disks.
Figure~\ref{snap} shows the phase diagram in the $(T,\mu)$-plane
together with several snapshots of typical vortex configurations
obtained in our simulations.  Below the KT transition line the system
is in a superfluid phase where all vortices present in the system are
bound together in pairs of opposite vorticity. In this phase the
vortex-vortex correlation function decays algebraically so there is no
LRO, but nevertheless the long distance macroscopic dielectric
function $\epsilon$ is finite, corresponding to a phase with finite
macroscopic superfluid density.  Above the KT transition, free
vortices appear and the macroscopic superfluidity is destroyed.  Our
simulation shows that this happens at any choice of the chemical
potential as long as it is negative.  In Fig.~\ref{snap}, the KT
transition line extends some (small) distance into the region of
positive chemical potential, and is then taken over by a first order
transition line at low temperature and small positive chemical
potential.

Our simulations are at finite temperature, but at zero temperature we
have the following three distinct ground states: an empty system (no
vortices) at low vortex chemical potential, a square vortex-antivortex
crystal above a certain chemical potential, which is always positive,
and a close-packed triangular (frustrated) vortex crystal above a
second threshold value of the chemical potential.  The crystal states
are indicated schematically in Fig.~\ref{snap}.  We simulated down to
low, but finite temperatures, but did not reach low enough to observe
any traces of these crystal states.  We can therefore not decide if
crystalline order, with finite dielectric response function, remains
at small but non-zero temperatures.  Long range crystal order can
also be destroyed at zero temperature by quantum melting, which is not
included in our classical model.  However, instead of crystalline LRO,
we find the typical short range positional correlations shown in the
inset in Fig.~\ref{snap}, close to the region of the first order
transition.  There are tendencies to form both 2D vortex-antivortex
finite clusters and also characteristic 1D vortex-antivortex strings
appear.  The configurations are strongly fluctuating, with strong
density fluctuations, and a snapshot at a later time will look very 
different but with the same kind of typical positional correlations.

Qualitatively different phase diagrams containing vortex-antivortex
crystal states have been discussed in the literature.  We do not
observe the thermally created vortex-antivortex crystals in
superfluids/superconductors that were proposed in
Refs~\cite{Kapitulnik,Zhang}.

Finally we will discuss the possible experimental consequences of our
phase diagram.  In the usual applications, like superfluid and
superconducting films, where the only vortices present are due to
thermal fluctuations (i.e.\ at negative vortex chemical potential),
our simulations show that there will always be a KT
transition.  According to our results, first order transitions can
happen only when the creation energy for a vortex pair is {\em
negative}, i.e., for positive chemical potential.  To enter this
parameter regime it is necessary to have a short range repulsion
between the vortices, such as in the hard disk model.  We expect that
this is not usually the case in superconducting or superfluid films,
but it should not be excluded that quantum fluctuations could
effectively lead to this situation in peculiar cases.  In any case our
simulation shows that a KT transition is not the only
possibility in this parameter regime.  An example of a physical
system, which has many of the properties of our CG model in the
parameter regime where we predict first order transitions, is provided
by spin textures in quantum Hall systems.  In double layer quantum
Hall systems at the filled lowest Landau level, i.e.\
one flux quantum of the perpendicular magnetic field penetrates the
sample for each electron in the 2D electron gas, a finite temperature
KT transition has been predicted~\cite{DLQHS}, and studied by lattice
simulations~\cite{Igor}.  Away in filling from the filled Landau level there
will be extra charges or holes present.  There is a certain regime
where the excess charge cause the formation of electrically charged
pseudospin vortices~\cite{DLQHS}, such that each vortex carries
electric charge $\pm e/2$.  This system thus automatically has a
finite density of vortex-antivortex pairs also at low temperature, so
at low enough temperature the vortex chemical potential is positive.
Furthermore, the hard disks correspond very roughly to the Coulomb
repulsion between the electric charges.  Thus the basic requirements
for observing our first order transitions are fulfilled, but clearly a
more detailed analysis taking into account for example the long range
nature of the electric Coulomb repulsion is needed.  This motivates
further simulations of this physical system using more realistic
models.

We thank Steve Girvin, Kenneth Holmlund, Edwin Langmann, Petter
Minnhagen, Peter Olsson, and Hans Weber for valuable discussions.
This work was supported by the Swedish Natural Science Research
Council.

\end{document}